\documentclass[prb,twocolumn,notitlepage,longbibliography]{revtex4-2}
\usepackage{amsmath}
\usepackage{amssymb}

\usepackage[unicode=true,colorlinks=true,citecolor=blue,urlcolor=blue]{hyperref}

\usepackage{bm}
\usepackage{epsfig}
 
\usepackage[normalem]{ulem}

\renewcommand{\phi}{\varphi}

\renewcommand {\Im}{\mathop\mathrm{Im}\nolimits}

\renewcommand {\phi}{{\varphi}}
\newcommand {\rmi}{{\rm i}}

\newcommand {\e}{{\rm e}}

\newcommand {\rot}{\mathop\mathrm{rot}\nolimits}
\begin{document}
\title{
Optomechanical circulator with a polaritonic microcavity
}

\author{Alexander V. Poshakinskiy}
\affiliation{Ioffe Institute, St. Petersburg 194021, Russia}

\author{Alexander N. Poddubny}
\email{poddubny@coherent.ioffe.ru}

\affiliation{Ioffe Institute, St. Petersburg 194021, Russia}

\begin{abstract}
We study theoretically optomechanical interactions in a semiconductor microcavity with embedded quantum well under the optical pumping  by a  Bessel beam, carrying a non-zero orbital momentum. Due to the transfer of  orbital momentum from light to phonons, the microcavity can act as an acoustic circulator: it rotates the propagation direction of the incident phonon by a certain angle clockwise or anticlockwise. Due to the optomechanical heating and cooling effects,  the circulator can also function as  an acoustic laser emitting sound with nonzero angular momentum. Our calculations demonstrate the potential of semiconductor microcavities for compact integrable optomechanical devices.
\end{abstract}
\date{\today}

\maketitle

\section{Introduction}\label{sec:intro}
Circulator is a multiport device  that  routes incoming signals directionally, either clockwise or counterclockwise.
Electromagnetic circulators operating at microwave and radiofrequencies have been realized since the 1950s using bulky magnetic setups~\cite{Chait1959}. However, it is an ongoing quest to implement  circulators that are compact~\cite{Mahoney2017} and potentially operating in the quantum regime~\cite{Kerckhoff2015}. The circulator effect implies nonreciprocal transmission and requires the breaking of the  time-reversal invariance \cite{Onsager1931,landau5}.  Apart from external magnetic field, the time-reversal symmetry can be also broken by coherent excitation of nonlinear systems or by time-modulation of structure paramters~\cite{Estep2014}. A very simple  mechanical non-magnetic realization of  acoustic circulator has been reported in Ref.~\cite{Fleury2014}, where  sound  passing through a ring resonator has been dragged clockwise by a rapidly spinning fan. This  concept can be potentially minituarized  by using active-liquid metamaterials as has been proposed in Ref.~\cite{Souslov2017}.
Even more opportunities to drive nonreciprocal effects in nanostructures are offered by optomechanical interactions that are inherently nonlinear~\cite{Hafezi2012,Peano2015,Poshakinskiy2017,Brendel2017}. Many experimental observations of optomechanical nonreciprocity are already available~\cite{Li2014c,Ruesink2016,Shen2018}, but the strength of optomechanical interactions and compatibility with the existing planar semiconductor technologies  can still be optimized.

\begin{figure}[b!]
\centering\includegraphics[width=.45\textwidth]{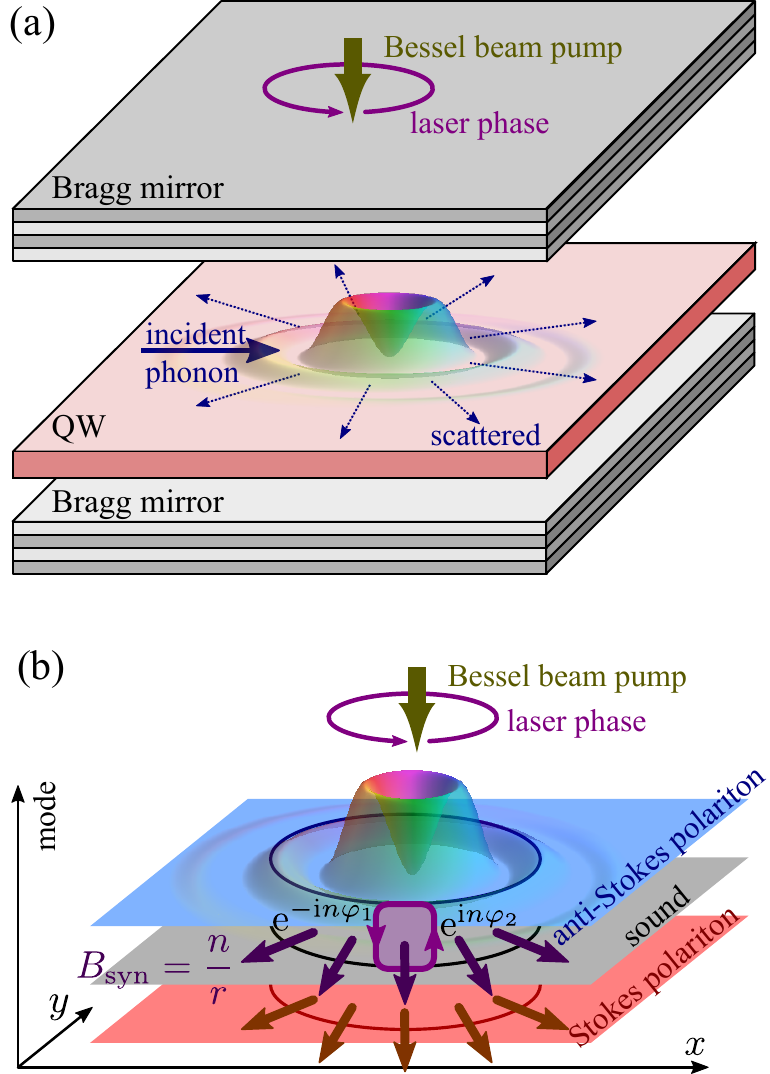}
\caption{(a) A sketch of the system: pillar microcavity made of a quantum well sandwiched between the Bragg  mirrors, that confines both light and sound and is pumped optically with a Bessel beam. 
(b) The synthetic magnetic field arising in the virtual space of three modes: sound, Stokes and anti-Stokes polariton. The phase acquired upon the indicated cycle of the pump-induced mode conversion is nonzero due to the lateral variation of the pump laser phase. The corresponding synthetic magnetic field resembles the field of a magnetic monopole~\cite{Dirac1931} with the charge equal to the order of the Bessel beam $n_L$.  This  synthetic magnetic field  affects propagation of coupled sound and polaritons in the plane of the microcavity leading to the circulator effect. 
} \label{fig:1}
\end{figure}

Here we theoretically propose a simple concept of an on-chip  acoustical circulator based on a quantum well embedded in the Bragg microcavity, that confines photons and acoustic waves simultaneously~\cite{Fainstein2017}, as illustrated in Fig.~\ref{fig:1}(a).  In such structures, the hybridization of quantum well excitons with light leads to formation of hybrid excitations, polaritons. Since both exciton and  photon part of the polariton interact with sound~\cite{Fainstein2013,Rozas2014}, the overall optomechanical interaction can be strongly increased~\cite{Jusserand2015}. For example, polariton-driven phonon laser operating in the regime of nonequilibrium Bose condensation of polaritons has recently been demonstrated~\cite{Chafatinos2020}. We consider a  microcavity illuminated from top by a Bessel beam that carries an orbital angular momentum $n_L$ . We study  the in-plane propagation of  mixed polariton-sound excitations in the presence of  such structured pump. Our calculations demonstrate that the Bessel beam of nonzero order induces a synthetic magnetic field in the virtual space spanned by phonons, Stokes and anti-Stokes polaritons, see Fig.~\ref{fig:1}(b). Indeed, the phase gained upon conversion of phonon to Stokes and anti-Stokes polariton is determined by the phase of pump pulse in the point of conversion~\cite{Poshakinskiy2017}. In the process of conversion forth and back in  different points, which corresponds to the round trip in virtual space indicated in Fig.~\ref{fig:1}(b), the total gained phase is nonzero. 
The corresponding  magnetic field is the field of the quantized magnetic monopole~\cite{Dirac1931} with the charge equal to the order of the Bessel beam $n_L$.  We show that this field leads to the circulator effect: propagating  waves are directionally scattered in the microcavity plane preferentially clockwise or counterclockwise depending on the sign of $n_L$.

The rest of the paper is organized as follows.  We start in Sec.~\ref{sec:B} by presenting the theoretical model for optically-induced synthetic magnetic field acting upon the sound. The calculated sound scattering amplitudes and the essence of the optomechanical circulator effect are discussed in Sec.~\ref{sec:circulator}. Section~\ref{sec:laser} analyses the role of  optomechanical  amplification effect for the circulator and we predict that this effect can drive a nonreciprocal light-induced circular acoustic laser. Section~\ref{sec:summary} summarizes the results and Appendices~\ref{sec:Appendix} and \ref{sec:Appendix2} are reserved for the details  of the numerical approach and for the perturbative scattering theory.

\section{Light-induced magnetic field for sound}\label{sec:B}
The structure under consideration is schematically illustrated in Fig.~\ref{fig:1}. It  is inspired by the micropillar cavities from
Refs.~\cite{Fainstein2017,Fainstein2013} and presents a semiconductor quantum well sandwiched between two distributed Bragg reflectors made from the pairs of alternating  GaAs and AlAs layers.  Interaction of semiconductor excitons inside the quantum well with photons trapped in the microcavity leads to formation of hybrid half-light half-matter excitations, excitonic polaritons~\cite{kavbamalas}.

The advantage of the pair of materials $\rm Ga{}_{0.9}Al{}_{0.1}As$ and $\rm Ga{}_{0.05}Al{}_{0.95}As$ is that it can act as a Bragg mirror simultaneously for light and sound due to the matching ratios of both sound and light velocities~\cite{Fainstein2013}. As such, the micropillar can trap both photons and phonons at the same time.
The system is pumped from above by a optical laser beam with the frequency $\omega_L$ that carries a non-zero orbital angular momentum $n_L$. In particular, we consider Bessel-Gauss beams 
\begin{equation}\label{eq:BG}
b_L(\bm r)= b_L J_{n_L}(qr)\, \e^{\rmi n_L \phi}\, \e^{-(r/r_0)^2},
\end{equation} 
where $r$ and $\varphi$ are the polar coordinates, $r_0$ is the width of the beam.
We  disregard the polarization effects due to the splitting of polaritons with transverse electric and magnetic polarization that is justified because the light propagation in the cavity is close to paraxial~\cite{kavbamalas}. The pump beam is scattered on phonons leading to the formation of Stokes and anti-Stokes polaritonic waves 
\[
b_{\rm S}(\bm r)\e^{-\rmi (\omega_L-\Omega)t},\quad b_{\rm aS}(\bm r)\e^{-\rmi (\omega_L+\Omega)t} \:,
\]
where $\Omega$ is the phonon frequency. Multiple Stokes and anti-Stokes scattering of polaritons, stimulated by the pump beam, leads to hybridization of polaritons with phonons, and formation of excitations termed as phonoritons~\cite{Hanke1999,Keldysh1986,Ivanov1982}. Such pump-induced interaction  is described by the following system of coupled equations~\cite{Poshakinskiy2017}
for the deformation amplitude $a(\bm r)$ and the polaritonic amplitudes $b_{\rm aS}(\bm r)$, $b_{\rm S}(\bm r)$:
\begin{align}
&\Omega a=\left(\Omega_{0}-\frac{\hbar\Delta}{2M}-\rmi \Gamma\right)a+g b_{L}^{*}(\bm r) b_{\rm aS}+g b_{L}(\bm r) b_{\rm S}^*\:,\label{eq:main}\\
&(\omega_L+\Omega) b_{\rm aS}=\left(\omega_{L}-\omega_{0}-\rmi \gamma-\frac{\hbar\Delta}{2m}\right) b_{\rm aS}+g b_{L}(\bm r)a\:,\nonumber\\
&(\omega_L-\Omega) b_{\rm S}^*=\left(\omega_{L}-\omega_{0}+\rmi \gamma-\frac{\hbar\Delta}{2m}\right) b_{\rm S}^*+g b_{L}^*(\bm r)a\:.\nonumber
\end{align}
We assume parabolic dispersion of phonons and polaritons  
\begin{equation} 
\Omega_{\bm k}=\Omega_{0}-\rmi \Gamma_x+\frac{\hbar k^{2}}{2M},\quad \omega_{\bm k}=\omega_{0}-\rmi \gamma+\frac{\hbar k^{2}}{2m}
\end{equation}
with the effective masses $M$ and $m$, respectively. The parameters $\Gamma$ and  $\gamma$ describe the  damping of phonons and polaritons, respectively. The phonon-polariton interaction parameter $g$ is contributed by both resonant photoelastic coupling and geometric coupling mechanisms~\cite{Fainstein2013,Baker2014}.

Our goal is to investigate the modification of the sound and polariton propagation due to the presence of the pump.
The scattering problem for the  plane acoustic wave, propagating in the cavity plane and interacting with the pump beam
can be readily solved numerically and will be discussed in Sec.~\ref{sec:circulator}. Before proceeding to the numerical results, it is however instructive to make a perturbative analysis of Eqs.~\eqref{eq:main} in order to get a general understanding of the effect of the orbital momentum transfer between pump light and phonons.

The simplest approximation is to neglect the  polariton dispersion in the second and third Eqs.~\eqref{eq:main}, express $b_{\rm aS}$ and $b_{\rm S}$ via $a(\bm r)$, and substitute them in the first Eq.~\eqref{eq:main}, which yields
\begin{multline}\label{eq:a1}
\left(\Omega_{0}-\rmi\Gamma-\hbar\frac{\Delta}{2M}-\Omega\right)a(\bm r)=a(\bm r) |b(\bm r)|^{2}\\\times\Bigl(\frac{g^{2}}{\omega_{0}-\omega_{L}-\Omega_{0}-\rmi \gamma}-\frac{g^{2}}{\omega_{L}-\omega_{0}-\rmi \gamma-\Omega_{0}}\Bigr)\:.
\end{multline}
The two terms in the right-hand side of Eq.~\eqref{eq:a1} describe the effect of light on sound. They lead to the optomechanical  spring, heating and cooling effects, i.e. the modification of phonon frequency and lifetime induced by light~\cite{Kippenberg2014}. These effects are well known for a zero-dimensional cavities, where light is confined and can not propagate. Here, however, we focus on the effects due to the propagation of  polaritons in the cavity plane. Specifically, treating the mass terms $\propto \hbar \Delta/(2m)$ in Eqs.~\eqref{eq:main} as a perturbation we obtain two more terms
\begin{multline*}
\frac{g^{2}}{2m(\omega_{0}-\rmi\gamma-\omega_{L}-\Omega)^{2}}b_{L}^{*}(\bm r) \Delta [b_{L}(\bm r) a(\bm r)]
\\+\frac{g^{2}}{2m(\omega_{L}-\omega_{0}-\rmi \gamma-\Omega)^{2}}b_{L}(\bm r) \Delta [b_{L}^{*}(\bm r) a(\bm r)]
\end{multline*}
that have to be added to the right hand side of Eq.~\eqref{eq:a1}. We are going to keep only the terms linear in $\nabla a$, since the remaining terms provide just corrections to the phonon mass and frequency.
The resulting equation for the phonon amplitude reads
\begin{equation}\label{eq:a-final}
\left(\Omega_{0}-\rmi\Gamma-\frac{p^{2}-2\bm A\cdot\bm p}{2M}+U\right) a=\Omega a,\: \bm p\equiv -\rmi \hbar \nabla\:.
\end{equation}
This is  an effective Schr\"odinger equation for phonons,  that experience light-induced scalar potential
\begin{equation}
U= -\frac{g^{2}|b(\bm r)|^{2}}{\omega_{0}-\rmi \gamma-\omega_{L}-\Omega_{0}}+\frac{g^{2}|b(\bm r)|^{2}}{\omega_{L}-\omega_{0}-\rmi \gamma-\Omega_{0}}\:,
\end{equation}
and light-induced vector potential
\begin{equation}
\bm A=\rmi \:u_{aS} b_{L}^{*} \nabla  b_{L}
+\rmi\:u_{S} b_{L} \nabla  b_{L}^{*}
\end{equation}
with
\begin{align*}\nonumber
u_{aS}&=\frac{Mg^{2}}{m(\omega_{0}-\rmi\gamma-\omega_{L}-\Omega_{0})^{2}},\\ 
u_{S}&=\frac{Mg^{2}}{m(\omega_{L}-\omega_{0}-\rmi\gamma-\Omega_{0})^{2}}\:.
\end{align*}
This vector potential corresponds to the synthetic magnetic field $\bm B=\rot\bm A$, directed perpendicular to the cavity plane,
\begin{multline}\label{eq:Bz}
B_z=\rmi u_{aS}\left[\frac{\partial b_{L}^{*}}{\partial x}\frac{\partial b_{L}}{\partial y}
-\frac{\partial b_{L}^{*}}{\partial y}\frac{\partial b_{L}}{\partial x}
\right]\\+
\rmi u_{S}\left[\frac{\partial b_{L}}{\partial x}\frac{\partial b_{L}^{*}}{\partial y}
-\frac{\partial b_{L}}{\partial y}\frac{\partial b_{L}^{*}}{\partial x}
\right]\:.
\end{multline}
Interestingly, the expression for the magnetic field is very similar to the Berry phase for the field $\bm b_L(\bm r)$, calculated in the real space~\cite{Berry1984}. The effective magnetic  field Eq.~\eqref{eq:Bz} appears  when the pump amplitude $\bm b_L$ carries nonzero orbital momentum. Specifically, since the Bessel beam  amplitude is proportional to $b_{L}\propto \e^{\rmi n_{L}\varphi}$ we find
\begin{equation}\label{eq:Ap}
\bm A\cdot\bm p= - \frac{\rmi n_{L}}{\rho^{2}}|b_{L}^{2}|(u_{aS}-u_{S})\frac{\partial}{\partial \phi}
\end{equation}
Hence, due to the term Eq.~\eqref{eq:Ap}, the propagation of sound becomes nonreciprocal. The clockwise and counterclockwise rotations are inequivalent due to the presence of the magnetic field $\bm B$.  More detailed analysis, performed in the following sections, will demonstrate that for the typical parameters of the semiconductor micropillars the  polariton dispersion term $\hbar \Delta/(2m)$ is not small. However, the qualitative conclusions   about the nonreciprocal phonon propagation remain valid. This is the essence of considered optomechanical circulator effect. 

\begin{figure}[t]
\includegraphics[width=.45\textwidth]{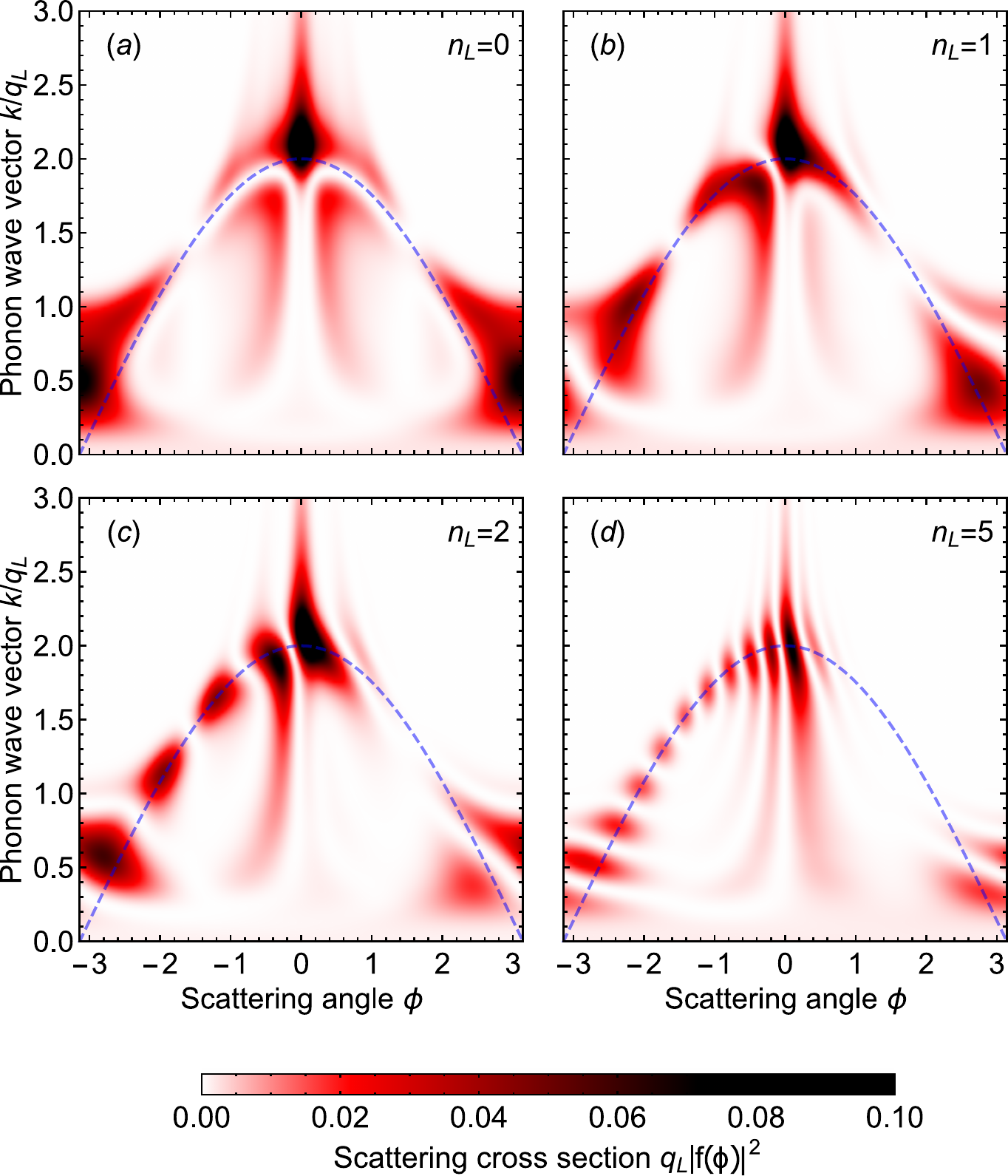}
\caption{The maps of phonon scattering probability as function of scattering angle and phonon wave vector.
The calculation is performed for the case of pumping with Bessel-Gauss beam, Eq.~\eqref{eq:BG}, with different angular momenta $n_L =0,1,2,5$, frequency detuning  $\omega_L-\omega_0 = \hbar^2q_L^2/(2m) = 1\,$meV and width $r_0 = 10/q_L$. The asymmetry of the map with respect to $\phi = 0$ that arises for $n_L \neq 0$ [panels (b)--(d)] indicates the scattering non-reciprocity. Dashed lines show the preferential scattering direction  calculated according to Eq.~\eqref{eq:theta}.
The structure parameters, taken from Ref.~\onlinecite{Fainstein2017}, are
$m  = 0.4\hbar^2\,\text{meV}^{-1}\cdot\mu\text{m}^{-2}$, $M =  10^4\hbar^2\,\text{meV}^{-1}\cdot\mu\text{m}^{-2}$, $\hbar\gamma = 0.1\,$meV, $2\pi\Omega_0  =19\,$GHz; the phonon damping $\Gamma$ is neglected. The optomechanical interaction strength is chosen to be $g|b_L| = 1\,\mu$eV, relevant for the maximal polariton density $|b_L|^2 = 10^9\,$cm$^{-2}$~\cite{Poshakinskiy2016PRL}. 
} \label{fig:map}
\end{figure}

\begin{figure}[t]
\includegraphics[width=.4\textwidth]{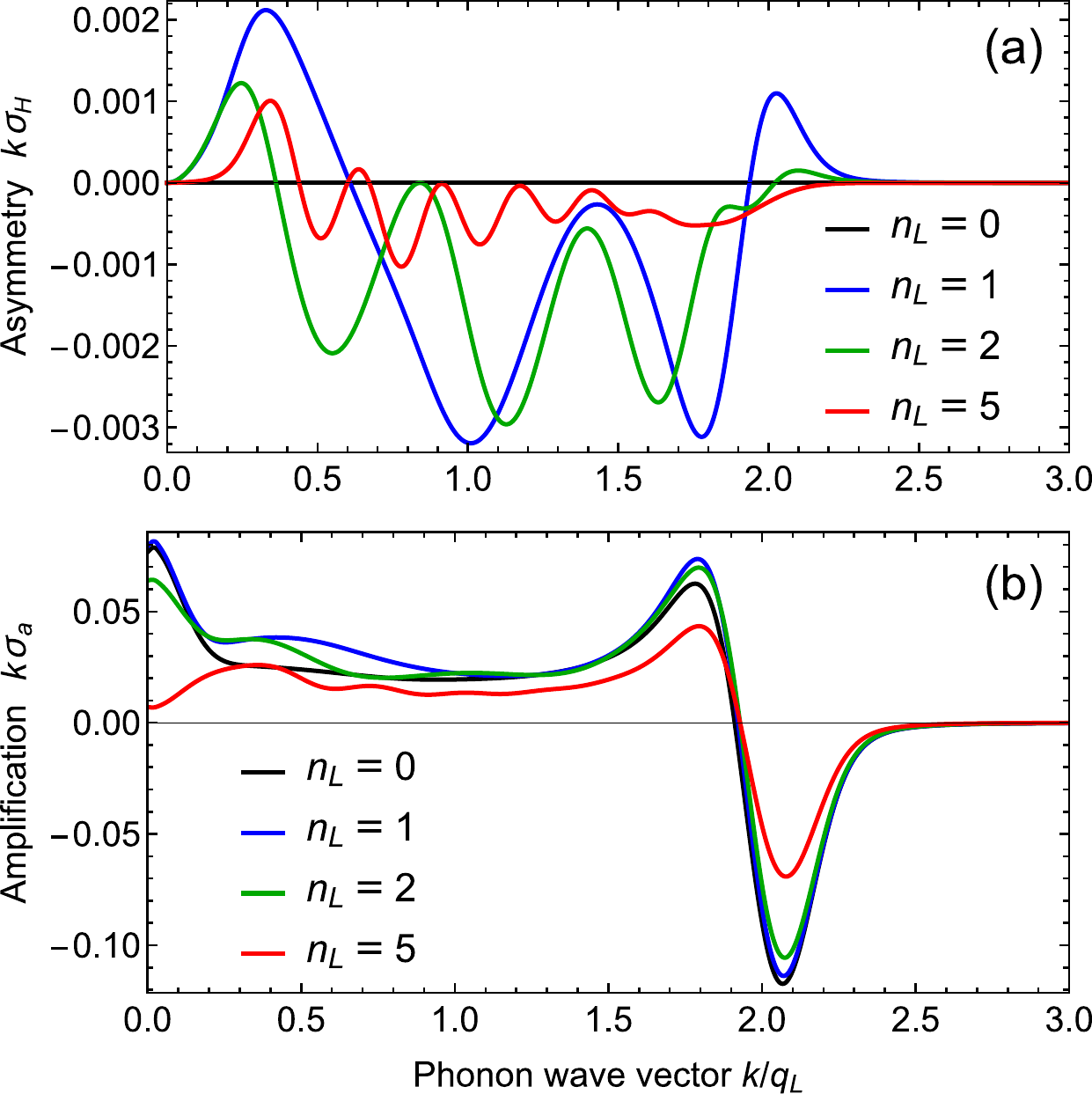}
\caption{(a) The transverse phonon current, characterizing the scattering asymmetry, and (b) phonon amplification calculated after Eqs.~\eqref{eq:sH},\eqref{eq:sa} as function of phonon wave vector.
The calculation is performed for the same parameters as Fig.~\ref{fig:map}. 
} \label{fig:AH}
\end{figure}

\section{Optomechanical circulator}\label{sec:circulator}
In this section we analyze the optomechanical  circulator effect in more detail. We consider the plane  acoustic wave with the wave vector $k$, that is incident along the $x$ direction and scatters on the Bessel pump beam. In the far-field region $\rho\gg 1/k$ the acoustic amplitude can be approximately written as 
\begin{equation}\label{eq:a-as}
a(r,\phi)=\e^{\rmi k r \cos \phi} + f(\phi)\,\frac{\e^{\rmi (kr - \pi/4)}}{\sqrt r}\:,
\end{equation}
where $f(\phi)$ is the scattering amplitude. The scattering cross section is given by $|f(\phi)|^2 d\phi$. The value of the scattering amplitude $f(\phi)$ can be readily calculated using the cylindrical symmetry of the problem. The numerical procedure is outlined in the Appendix~\ref{sec:Appendix} and the result assumes the form
\begin{align}\label{eq:f}
f(\phi) = \frac{1}{\sqrt{2\pi k}} \sum\limits_{n=-\infty}^\infty  [S_{11}(n)-1] \e^{\rmi n \phi}\:,
\end{align}
where $S_{11}(n)$ is the element of the scattering matrix~\eqref{eq:S} that describes reflection of the cylindrical acoustic wave with angular momentum $n$.

The dependence of the scattering pattern on the phonon wave vector $k$ and on the pump beam orbital momentum $n_L$ is shown in Fig.~\ref{fig:map}. The calculation has been performed for the realistic parameters, relevant for the experimental setup of  Ref.~\onlinecite{Fainstein2017}, see the  caption of Fig.~\ref{fig:map}. Different panels in Fig.~\ref{fig:map} correspond to different values of the light orbital momentum $n_L=0,1,2, 5$.  When the light orbital momentum increases, the scattering becomes  strongly directional. The preferential scattering angle is approximately given by the resonant condition 
\begin{equation}\label{eq:theta}
\cos\frac{\phi}{2}=\frac{k}{2q_L}\:,
\end{equation}
and shown by the dashed lines in Fig.~\ref{fig:map}. This condition follows from the perturbative treatment of the scattering discussed in Appendix~\ref{sec:Appendix2}.
Moreover, the calculation demonstrates that the scattering at angles $\varphi$ and $-\varphi$ becomes  asymmetric if the pump beam carries nonzero orbital momentum $n_L\ne 0$.  
In this case, due to the light-induced magnetic field Eq.~\eqref{eq:Bz}, the reflection of cylindrical acoustic waves with the opposite values of $n$ becomes different resulting in preferential clockwise (for $n_L>0$) or counterclockwise (for $n_L<0$) scattering of phonons. 

The asymmetry of the scattering map can be quantified by the component of scattered phonon current that is perpendicular to the propagation direction of the incident phonon. Such transverse current is similar to Hall conductivity in electronic gas and is described by
\begin{equation}\label{eq:sH}
\sigma_{\rm H} = \int \sin \phi \, |f(\phi)|^2\, d\phi \:.
\end{equation}
Figure~\ref{fig:AH}(a) shows the dependence of $\sigma_{\rm H}$ on the phonon wave vector. In case of excitation by the beam without angular momentum, $n_L=0$, transverse phonon current vanishes,  $\sigma_{\rm H} = 0$. The strongest transverse current can be achieved for $n_L=1$. 

Due to the optomechanical heating and cooling effects, the acoustic wave can be amplified or attenuated upon scattering. This effect can  characterized by the   amplification cross section
\begin{align}\label{eq:sa}
\sigma_a = \sqrt{\frac{8\pi}{k}} \text{Re\,} f(0)  + \int |f(\phi)|^2\, d\phi\:,
\end{align}
where the first term corresponds to the forward scattering while the second term describes scattering to all other directions. Figure~\ref{fig:AH}(b) shows the dependence of $\sigma_{\rm a}$ on the phonon wave vector. The strongest amplification and attenuation is achieved in the vicinity of resonant condition $k = 2 q_L$, corresponding to the scattering angle $\varphi$ close to zero, see Eq.~\eqref{eq:theta}.


\section{Nonreciprocal circular acoustic laser}\label{sec:laser}
In the previous section, we have analyzed the scattering problem when the external Bessel-beam pump leads to directional nonreciprocal scattering of propagating acoustic waves.  Here, we  study an eigenvalue problem for coupled polaritons and sound waves in the presence of the Bessel-beam pump. Our goal is to demonstrate that the combination of optomechanical heating effect and non-zero orbital momentum of pump light can lead to nonreciprocal acoustic laser, where the gain is realized only for the mode propagating clockwise (or only for the mode propagating counterclockwise).

To this end we make use of the cylindrical symmetry of the problem and substitute the solutions in the form
\begin{align}\label{eq:cyl}
& b_L (\bm r)  = b_L(r) \e^{\rmi n_L \phi}\:, \\\nonumber
& a (\bm r)  = a(r) \e^{\rmi n \phi}\:, \\\nonumber
& b_{\rm aS} (\bm r)  = b_{\rm aS}(r) \e^{\rmi (n_L+n) \phi}\:, \\\nonumber
& b_{\rm S} (\bm r)  = b_{\rm S}(r) \e^{\rmi (n_L-n) \phi}\:,
\end{align}
into the system Eq.~\eqref{eq:main}.

\begin{equation}\label{eq:an}
\end{equation}
\begin{figure}[t!]
\includegraphics[width=.45\textwidth]{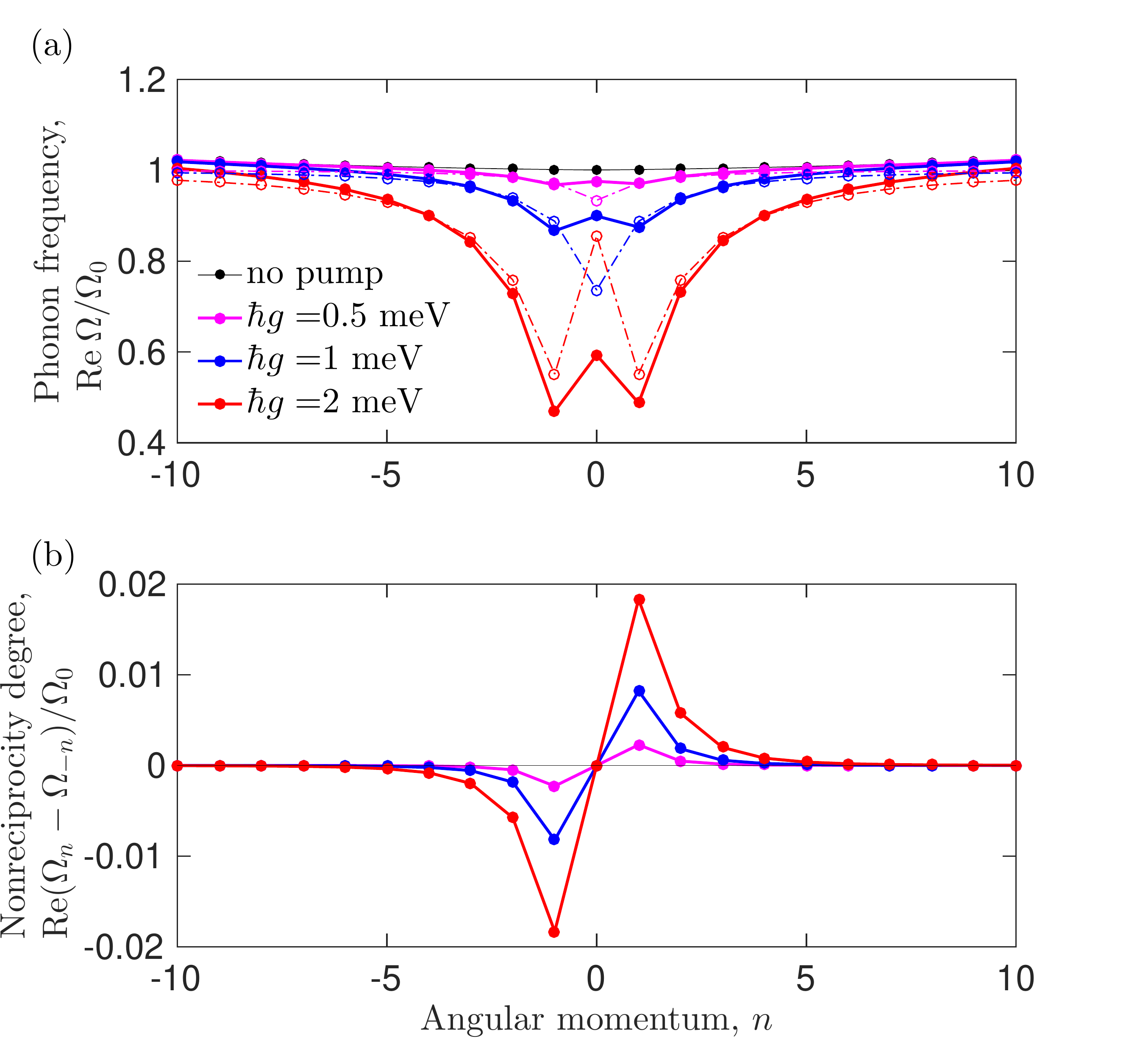}
\caption{
(a) Real parts of the phonon energy $\Omega_n$ of the fundamental mode and (b) the asymmetry of the phonon spectrum $\Omega_n-\Omega_{-n}$ calculated depending on the azimuthal momentum $n$ for different pump strengths $\hbar g$, indicated on the graph. 
Solid symbols correspond to the numerical calculation after Eq.~\eqref{eq:an}, open symbols present the solution of the semi-analytical equation \eqref{eq:mass}.
The calculation is performed for the  Bessel-Gauss beam, Eq.~\eqref{eq:BG}, with the angular momentum $n_L =1$,
$q_L=0.1$~$\mu$m${}^{-1}$, $r_0=2~\mu$m,  the cavity radius is $R=3~\mu$m, the pump frequency is detuned negatively with respect to polariton band edge, $\hbar(\omega_L-\omega_0)=-0.2$~meV, and phonon frequency is $\hbar\Omega=0.05$~meV. Other structure parameters are the same as in Fig.~\ref{fig:map}.
} \label{fig:3}
\end{figure}
The results of calculation are presented in Fig.~\ref{fig:3}.  We considered  the pillar of the finite radius $R=3\mu$m.
In the absence of the pump the frequency of the lowest quantized acoustic mode  is given by
\begin{equation}
\Omega=\Omega_0+\frac{\hbar \gamma_n^2}{2MR^2}\:,
\end{equation}
where $\gamma_n$ is the first zero of the $n$-th Bessel function. The corresponding dispersion curve is shown by black circles in Fig.~\ref{fig:3}(a). Due to the large value of the phonon mass, $M =  10^4\hbar^2\,\text{meV}^{-1}\cdot\mu\text{m}^{-2}\gg m$, their dispersion is quite weak and can be neglected. However, the phonon modes acquire dispersion  when the structure is illuminated by light. In Fig.~\ref{fig:3}, we consider the case of negative detuning, $\hbar(\omega_L-\omega_0)=-0.2$~meV. Then, the illumination leads mainly to the modification of the phonon frequency, i.e., the optical spring effect~\cite{Kippenberg2014}. The calculation demonstrates that the phonon frequency decreases for large pump strengths and the effective mass of the phonon mode becomes smaller. In order to describe this analytically we first rewrite  Eq.~\eqref{eq:main} as
\begin{multline}\label{eq:a2}
\left(\Omega_{0}-\rmi\Gamma-\hbar\frac{\Delta}{2M}-\Omega\right)a(\bm r)=b_L(\bm r)\\\times\Big(\frac{g^{2}}{\omega_{0}-\omega_{L}-\Omega_{0}+\frac{\hbar\Delta}{2m}-\rmi \gamma}\\-\frac{g^{2}}{\omega_{L}-\omega_{0}-\rmi \gamma-\Omega_{0}-\frac{\hbar\Delta}{2m}}\Big)b_L(\bm r)  a(\bm r) \:.
\end{multline}
The considered here regime of parameters is different from those in Sec.~\ref{sec:B}. Namely, the largest term in the denominators of Eq.~\eqref{eq:a2} is the one describing the polariton dispersion,
 \begin{equation}
 \frac{\hbar\Delta}{2m}\sim \frac{\hbar}{mR^2}\gg \gamma ,|\omega_{L}-\omega_{0}-\Omega_0|\:,\label{eq:regime}
 \end{equation}
 that can not be treated perturbatively.
Leaving only this term we obtain, the phonon dispersion can be approximately described by the equation
\begin{equation}\label{eq:mass}
(\Omega_0-\Omega-\rmi \Gamma)\Delta a=-4m |b_L|^2|g|^2a\:.
\end{equation}
Solution of Eq.~\eqref{eq:mass} is shown in Fig.~\ref{fig:3}(a) by open symbols and satisfactorily describes the results of full numerical calculation.

\begin{figure}[t!]
\includegraphics[width=.45\textwidth]{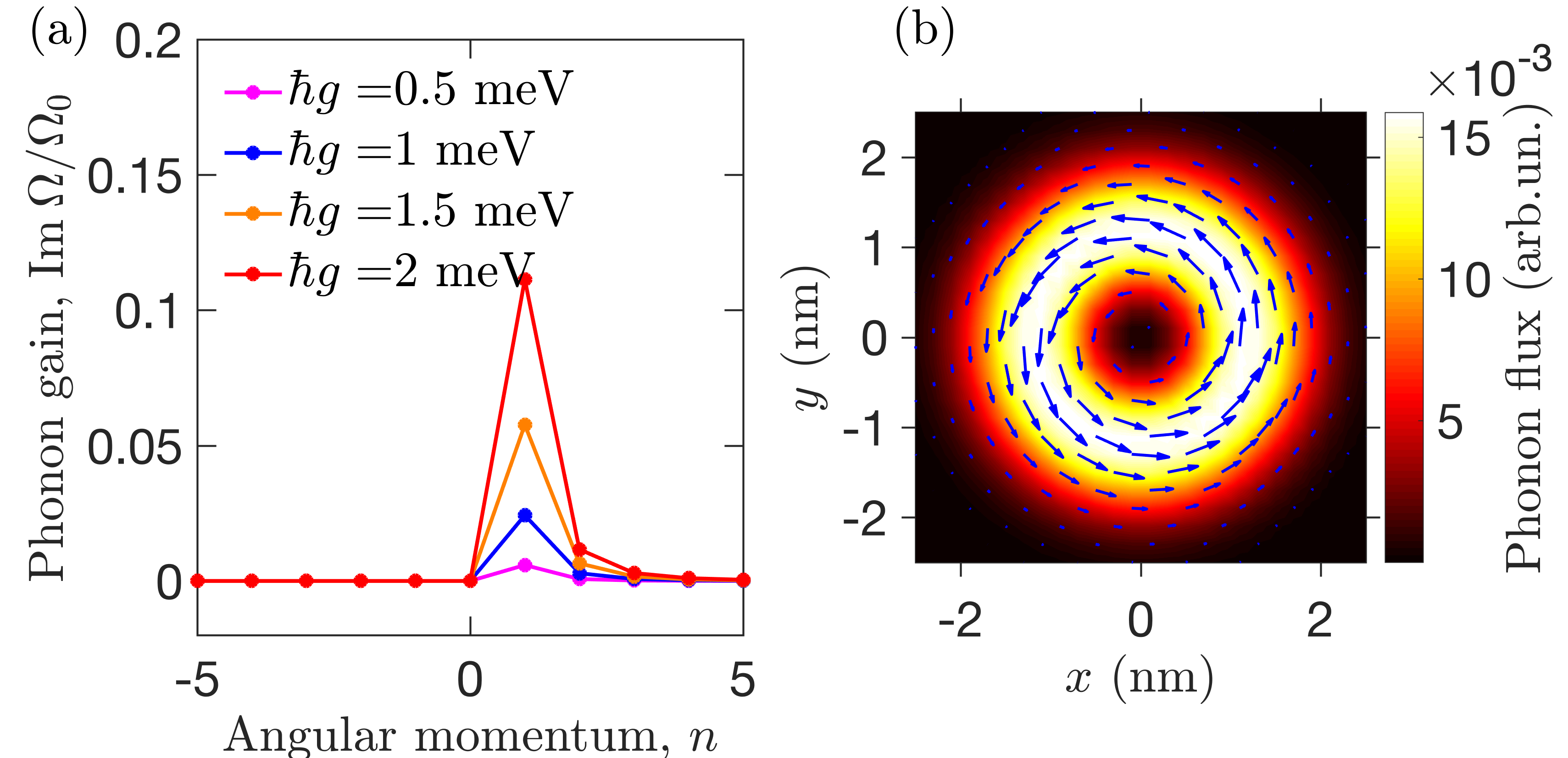}
\caption{(a) The gain of phonon modes $\text{Im\,}\Omega_n$ calculated depending on the azimuthal momentum $n$
for different pump strengths $\hbar g$, indicated on the graph.  Calculation has been performed for the pump frequency detuned positively with respect to the polariton band edge, $\hbar(\omega_L-\omega_0)=+0.2$~meV, and the same other parameters as in Fig.~\ref{fig:3}.
(b) Phonon flux calculated for the mode with maximum gain and azimuthal momentum $n=1$ for $\hbar g=2$~meV.
} \label{fig:4}
\end{figure}

 Crucially, the illumination not only changes the phonon mass and frequency, but also makes the phonon dispersion asymmetric,  $\Omega_n\ne \Omega_{-n}$, as can be seen from the lower panel of Fig.~\ref{fig:3}(b). 
The degree of asymmetry increases proportional to the pump intensity, $\Omega_n-\Omega_{-n}\propto g^2$.
This is in qualitative agreement with the analytical results in Sec.~\ref{sec:B} and can be viewed as a manifestation of light-induced synthetic magnetic field.  

We now proceed to analyze the regime of positive detuning, $\omega_L-\omega_0=0.2$~meV. In this case the phonon emission processes become more efficient than phonon absorption processes, which corresponds to the regime of phonon laser due to the optomechanical heating effect~\cite{Kippenberg2014}. Formally, the acoustic gain is characterized by positive imaginary part of the eigenfrequency, $\Im \Omega_n>0$.  The dependence of $\Omega_n$ on the azimuthal number $n$ is plotted for different pumping strength in Fig.~\ref{fig:4}(a). Similarly to the case of Fig.~\ref{fig:3}, the gain is  not a symmetric function of the azimuthal number $n$,
$\Im \Omega_n\ne \Im \Omega_{-n}$.
This is the regime of nonreciprocal lasing~\cite{Bahari2017,Segev2018b}. 
 This spatial profile of the mode  with maximum gain for $\hbar g=2$~meV and $n=1$ is illustrated in 
Fig.~\ref{fig:4}(b). Arrows show the magnitude and the direction of the acoustic flux $\bm j\propto \Im a^*(\bm r)\nabla \bm a(r)$. 
Clearly,  the lasing mode propagates in the counterclockwise direction dictated by the pump beam $n=n_L=1$.

\section{Summary}\label{sec:summary}
To summarize, we have studied the propagation of sound in the planar semiconductor microcavity that is optically pumped by a Bessel beam. We have demonstrated, that the optical pump can not only modify the phonon frequency, decay rate and effective mass, but also can induce an effective magnetic field for the propagating phonons.  As a result, the calculated  phonon scattering pattern induced by the pump beam becomes strongly asymmetric, realizing the regime of  optically-controlled acoustic circulator. In addition to the scattering problem, we have also analyzed the acoustic eigenstates of effective non-Hermitian potential induced by light for phonons. We demonstrated that the optomechanical heating effect leads to the amplification of acoustic modes, when can be used to create non-reciprocal phonon laser. Thus, the recently developed high quality semiconductor microcavities confining both light and sound~\cite{Fainstein2017} can be potentially used to route and amplify sound by light in a compact integrable solid-state setup. One can also envisage more complex phase and intensity profiles of the pump beam leading to the 
creation of topologically nontrivial phases of sound~\cite{Peano2015} that could be used to guide sound safely from disorder.

\begin{acknowledgments}
This work has been supported by the Russian Science Foundation Grant No.~20-42-04404.
\end{acknowledgments}

\appendix
\section{Numerical solution of the scattering problem}\label{sec:Appendix}
In this Appendix, we outline the approach to describe the scattering of phonons interacting with  the pump laser beam.
The numerical procedure follows the phase formalism from Ref.~\onlinecite{babikov1966method}, see also the Supplementary Information of Ref.~\onlinecite{Denisov2017}.
We assume that the pump is centrosymmetric, $b_L(\bm r) = b_L(r) \e^{\rmi n_L \phi}$ and  search for solution in the form Eq.~\eqref{eq:cyl}.
 We start from the ansatz
\begin{align}\label{eq:Babikov}
\mathbf v(r) = \left( \begin{array}{c} a(r) \\ b_{\rm aS}(r) \\b_{\rm S}^*(r) \end{array} \right) = [\hat J(r) + \hat N(r) \hat K(r)] \mathbf v_0(r) 
\end{align}
where $\hat J(r)$ and $\hat N(r)$ are the diagonal matrices of free solutions of the system \eqref{eq:main},
\begin{multline}\label{eq:J}
\hat J(r) = {\rm diag\,} [ \sqrt{2M} J_n(kr) , \sqrt{2m} J_{n+n_L}(q_{\rm aS}r) ,\\  \sqrt{2m} J_{n_L-n}(q_{\rm S}r)  ] \,,
\end{multline}
the $\hat N$ matrix is constructed the same way from the Neumann functions, and $q_{\rm aS(S)} = \sqrt{2m(\omega_L\pm \Omega - \omega_x \pm \rmi\Gamma)}$. The vector function $\mathbf v_0(r)$ and matrix function  $\hat K(r)$ are the parameters of the ansatz.
Without the loss of generality  we can set 
\begin{align}
\mathbf v'(r) =  \hat k [\hat J'(r) + \hat N'(r) \hat K(r)] \mathbf v_0(r)\,,
\end{align}
where the prime means differentiation over $r$ and $\hat k=  {\rm diag\,} ( k,q_{\rm aS},q_{\rm S} )$. 
This leads to the condition 
\begin{align}
\mathbf v_0' =- [ \hat J+  \hat N \hat K]^{-1} \hat N \hat K' \mathbf v_0 \,.
\end{align}
Substituting the ansatz Eq.~\eqref{eq:Babikov} to the initial scattering Eq.~\eqref{eq:main} we obtain
\begin{align}\label{eq:dK}
\hat K' = (\hat J + \hat K \hat N) \frac{2\hat m}{ k \hat W} \hat V (\hat J +  \hat N \hat K)\:.
\end{align}
The initial condition $\hat K(0) = 0$ is additionally imposed to make the solutions regular at the origin. Here, $\hat W = \hat J \hat N' - \hat J' \hat N$ is the Wronskian that for the free solutions in the form of Eq.~\eqref{eq:J} reads
\begin{align}
\hat W = 2\hat m \, \frac{2}{\pi \hat k r}
\end{align}
and 
\begin{align}
\hat V = \left( \begin{array}{ccc} 0 & g b_L^* & g b_L \\ g b_L & 0 &0\\ g b_L^* &0& 0\end{array} \right)
\end{align}
is the matrix of optomechanic coupling from equation set~\eqref{eq:main}.
In a little bit more explicit way we rewrite Eq.~\eqref{eq:dK} as 
\begin{align}\label{eq:dK2}
\hat K' = (\hat J + \hat K \hat N) \frac{\pi r}{2} \hat V (\hat J +  \hat N \hat K)\:.
\end{align}
For each value of $n$, Eq.~\eqref{eq:dK2} is just a first-order linear matrix differential equation that can be readily solved numerically to find the matrix $ \hat K(\infty)$, that determines the scattered waves. 
Specifically, in the asymptotic region of $r\to \infty$ the field can be presented as
\begin{equation}\label{eq:asymp}
\mathbf v\propto [(J-\rmi N \Sigma)+(J+\rmi N \Sigma )S]\mathbf v_{0}\:,
\end{equation}
where the matrix $\hat \Sigma = {\rm diag\,} (1,1,-1)$ accounts for the Bogolyubov nature of excitations.
Comparing Eq.~\eqref{eq:asymp} with Eq.~\eqref{eq:Babikov},
we find the scattering matrix
\begin{align}\label{eq:S}
\hat S = \frac{1-\rmi\hat\Sigma \hat K(\infty)}{1+\rmi\hat\Sigma \hat K(\infty)}\:.
\end{align}
We note   that $\hat S^\dag \hat\Sigma \hat S =1$ in the absence of losses.
\subsection*{Scattering cross sections}
Here we present useful expressions for different cross sections characterizing the scattering process.
Using the asymptotic expression Eq.~\eqref{eq:a-as}  for the sound amplitude and the 
 the plane wave  decomposition
\begin{align}
\e^{\rmi k r \cos \phi} = \sum_{n=-\infty}^{+\infty} \rmi^n J_n(kr) \,\e^{\rmi n \phi}
\end{align}
we obtain the scattering amplitude 
\begin{align}
f(\phi) = \frac{1}{\sqrt{2\pi k}} \sum_n [S_{11}(n)-1]\, \e^{\rmi n \phi}\:
\end{align}
(see also Eq.~\eqref{eq:f} in the main text), where $S_{11}(n)$ means the component of the scattering matrix with the azimuthal momentum $n$, corresponding to the phonon scattering (index $1$).
The total scattering cross section is then given by
\begin{align}
\sigma_s = \int |f(\phi)|^2\, d\phi = \frac1{k}\sum_n |S_{11}(n)-1|^2\:.
\end{align}
The flux of the  forward going acoustic wave reads
\begin{align}
\sigma_0 = \sqrt{\frac{2\pi}{k}} \left[ f(0) + f^*(0)\right] = \frac1k \sum_n \left[ S_{11}(n)+S_{11}^*(n)-2\right]\:.
\end{align}
The acoustic wave can be amplified or attenuated upon the scattering, $|S_{11}(n)| \neq 1$. This effect is  characterized by the   amplification cross section
\begin{align}
\sigma_a = \sigma_s + \sigma_0 =  \frac1{k}\sum_n \left[ |S_{11}(n)|^2-1 \right]\:.
\end{align}

The scattering asymmetry, essential for the optomechanical circulator effect, can described by the ``Hall'' cross-section.
It determines appearance of the transversal phonon current and reads
\begin{equation}
\sigma_{\rm H} = \int \sin \phi \, |f(\phi)|^2\, d\phi = \frac1k \,{\rm Im\, }\sum_n S_{11}(n)\,S_{11}^*(n+1)\:.
\end{equation}
We note that the Hall cross section $\sigma_{\rm H}$ vanishes at $n_L=0$ when $S(n)=S(-n)$.


\section{Perturbative approach for the scattering}\label{sec:Appendix2}
\begin{figure}[t!]
\centering\includegraphics[width=0.2\textwidth]{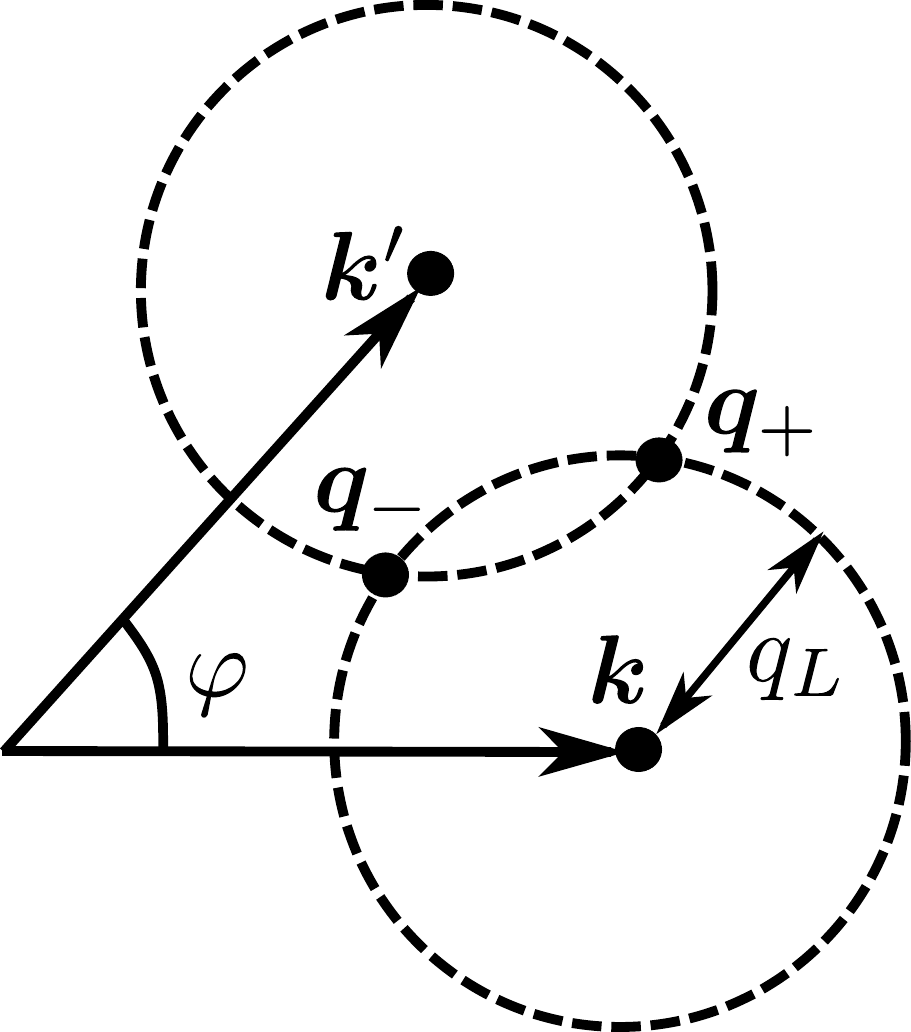}
\caption{Illustration of the momentum conservation laws for the scattering. Wave vectors $\bm k$ and $\bm k'$ correspond to the initial and final states of the phonons, wave vectors $\bm q_\pm$ describe intermediate virtual states of the polaritons.}\label{fig:qk}
\end{figure}
In this Appendix, we present a simplified approach to calculate the phonon scattering amplitude in the second order in optomechanical coupling constant $g$. The calculation is conveniently performed in the reciprocal space. 
For the considered case of  the excitation with Bessel beam
$
b_L(\bm r) = b_L J_{n_L}(qr)\, \e^{\rmi n_L \phi}$, 
the Fourier transform of the excitation amplitude reads
\begin{equation}
b_L(\bm q)  =  \frac{2\pi (-\rmi)^{n_L} b_L}{q_L}  \delta(q-q_L) \,\e^{\rmi n_L \phi}\:.
\end{equation}
The phonon scattering amplitudes due to conversion to anti-Stokes and Stokes polaritons in the leading order of perturbation theory are then given by
\begin{align}
&T_{\bm k' \bm k}^{\rm (aS)} = -\rmi g^2 \sum_{\bm q} b_L(\bm q-\bm k) b_L^*(\bm q-\bm k') \frac{1}{\omega_L+\Omega_k-\omega_q+\rmi\Gamma} \\
&T_{\bm k' \bm k}^{\rm (S)} = -\rmi g^2 \sum_{\bm q} b_L(\bm q+\bm k') b_L^*(\bm q+\bm k) \frac{1}{\omega_L-\Omega_k-\omega_q-\rmi\Gamma}\:.\nonumber
\end{align}
The integral over $\bm q$ is determined by the two points with the  wave vectors of the intermediate polariton state 
\begin{align}
q_\pm = k \cos\frac\phi2 \pm  \sqrt{q_L^2-k^2  \sin^2 \frac\phi2}\:,
\end{align}
given by the intersection of two circles with the radius $q_L$ centered at the points $\bm k$ and $\bm k'$, as show in Fig.~\ref{fig:qk} .

  After the integration over $\bm q$ we obtain the  following amplitude for scattering by an angle $\phi$:
\begin{multline}\label{eq:T}
T = -\frac{\rmi g^2}{2 k \left| \sin \frac\theta2 \right| \, \sqrt{q_L^2-k^2  \sin^2 \frac\phi2}}\\\times \sum_{s,s'=\pm} \frac{\e^{2 \rmi ss' n \arcsin\left(\frac{k}{q_L} \sin \frac{\phi}2 \right) }}{\omega_L -\omega_{q_s} + s'(\Omega_k+\rmi\Gamma)}\:,
\end{multline}
where 
\begin{align}
\omega_{q_\pm} - \omega_L = \frac{k^2 {\cos^2 \phi}}{2m} \pm \frac km \cos\frac\phi2   \sqrt{q_L^2-k^2  \sin^2 \frac\phi2}\:.
\end{align}
In the considered order, the scattering amplitude is
\begin{align}
f = \frac{M}{\sqrt{2\pi k}} \, T
\end{align}
and the amplification cross-section reads
\begin{align}
\sigma_a = \frac{M}{k}\, 2\,{\rm Re\,}T(0) \:.
\end{align}
Analysis of the resonant denominators in the second line of Eq.~\eqref{eq:T} shows that the strongest scattering corresponds to the case when the intermediate polariton wave vector $q_-$ is equal to the laser wave vector, $q_-=q_L$, which is realized at
 \begin{equation}\label{eq:theta1}
k=2q\cos\frac{\phi}{2}\:,
\end{equation}
or Eq.~\eqref{eq:theta}  from the main text.


\begin{thebibliography}{33}%
\makeatletter
\providecommand \@ifxundefined [1]{%
 \@ifx{#1\undefined}
}%
\providecommand \@ifnum [1]{%
 \ifnum #1\expandafter \@firstoftwo
 \else \expandafter \@secondoftwo
 \fi
}%
\providecommand \@ifx [1]{%
 \ifx #1\expandafter \@firstoftwo
 \else \expandafter \@secondoftwo
 \fi
}%
\providecommand \natexlab [1]{#1}%
\providecommand \enquote  [1]{``#1''}%
\providecommand \bibnamefont  [1]{#1}%
\providecommand \bibfnamefont [1]{#1}%
\providecommand \citenamefont [1]{#1}%
\providecommand \href@noop [0]{\@secondoftwo}%
\providecommand \href [0]{\begingroup \@sanitize@url \@href}%
\providecommand \@href[1]{\@@startlink{#1}\@@href}%
\providecommand \@@href[1]{\endgroup#1\@@endlink}%
\providecommand \@sanitize@url [0]{\catcode `\\12\catcode `\$12\catcode
  `\&12\catcode `\#12\catcode `\^12\catcode `\_12\catcode `\%12\relax}%
\providecommand \@@startlink[1]{}%
\providecommand \@@endlink[0]{}%
\providecommand \url  [0]{\begingroup\@sanitize@url \@url }%
\providecommand \@url [1]{\endgroup\@href {#1}{\urlprefix }}%
\providecommand \urlprefix  [0]{URL }%
\providecommand \Eprint [0]{\href }%
\providecommand \doibase [0]{http://dx.doi.org/}%
\providecommand \selectlanguage [0]{\@gobble}%
\providecommand \bibinfo  [0]{\@secondoftwo}%
\providecommand \bibfield  [0]{\@secondoftwo}%
\providecommand \translation [1]{[#1]}%
\providecommand \BibitemOpen [0]{}%
\providecommand \bibitemStop [0]{}%
\providecommand \bibitemNoStop [0]{.\EOS\space}%
\providecommand \EOS [0]{\spacefactor3000\relax}%
\providecommand \BibitemShut  [1]{\csname bibitem#1\endcsname}%
\let\auto@bib@innerbib\@empty
\bibitem [{\citenamefont {{Chait}}\ and\ \citenamefont
  {{Curry}}(1959)}]{Chait1959}%
  \BibitemOpen
  \bibfield  {author} {\bibinfo {author} {\bibfnamefont {H.~N.}\ \bibnamefont
  {{Chait}}}\ and\ \bibinfo {author} {\bibfnamefont {T.~R.}\ \bibnamefont
  {{Curry}}},\ }\bibfield  {title} {\enquote {\bibinfo {title} {{Y
  Circulator}},}\ }\href {\doibase 10.1063/1.2185863} {\bibfield  {journal}
  {\bibinfo  {journal} {J. Appl. Phys.}\ }\textbf {\bibinfo {volume} {30}},\
  \bibinfo {pages} {S152--S153} (\bibinfo {year} {1959})}\BibitemShut {NoStop}%
\bibitem [{\citenamefont {Mahoney}\ \emph {et~al.}(2017)\citenamefont
  {Mahoney}, \citenamefont {Colless}, \citenamefont {Pauka}, \citenamefont
  {Hornibrook}, \citenamefont {Watson}, \citenamefont {Gardner}, \citenamefont
  {Manfra}, \citenamefont {Doherty},\ and\ \citenamefont
  {Reilly}}]{Mahoney2017}%
  \BibitemOpen
  \bibfield  {author} {\bibinfo {author} {\bibfnamefont {A.~C.}\ \bibnamefont
  {Mahoney}}, \bibinfo {author} {\bibfnamefont {J.~I.}\ \bibnamefont
  {Colless}}, \bibinfo {author} {\bibfnamefont {S.~J.}\ \bibnamefont {Pauka}},
  \bibinfo {author} {\bibfnamefont {J.~M.}\ \bibnamefont {Hornibrook}},
  \bibinfo {author} {\bibfnamefont {J.~D.}\ \bibnamefont {Watson}}, \bibinfo
  {author} {\bibfnamefont {G.~C.}\ \bibnamefont {Gardner}}, \bibinfo {author}
  {\bibfnamefont {M.~J.}\ \bibnamefont {Manfra}}, \bibinfo {author}
  {\bibfnamefont {A.~C.}\ \bibnamefont {Doherty}}, \ and\ \bibinfo {author}
  {\bibfnamefont {D.~J.}\ \bibnamefont {Reilly}},\ }\bibfield  {title}
  {\enquote {\bibinfo {title} {On-chip microwave quantum hall circulator},}\
  }\href {\doibase 10.1103/PhysRevX.7.011007} {\bibfield  {journal} {\bibinfo
  {journal} {Phys. Rev. X}\ }\textbf {\bibinfo {volume} {7}},\ \bibinfo {pages}
  {011007} (\bibinfo {year} {2017})}\BibitemShut {NoStop}%
\bibitem [{\citenamefont {Kerckhoff}\ \emph {et~al.}(2015)\citenamefont
  {Kerckhoff}, \citenamefont {Lalumi\`ere}, \citenamefont {Chapman},
  \citenamefont {Blais},\ and\ \citenamefont {Lehnert}}]{Kerckhoff2015}%
  \BibitemOpen
  \bibfield  {author} {\bibinfo {author} {\bibfnamefont {J.}~\bibnamefont
  {Kerckhoff}}, \bibinfo {author} {\bibfnamefont {K.}~\bibnamefont
  {Lalumi\`ere}}, \bibinfo {author} {\bibfnamefont {B.~J.}\ \bibnamefont
  {Chapman}}, \bibinfo {author} {\bibfnamefont {A.}~\bibnamefont {Blais}}, \
  and\ \bibinfo {author} {\bibfnamefont {K.~W.}\ \bibnamefont {Lehnert}},\
  }\bibfield  {title} {\enquote {\bibinfo {title} {On-chip superconducting
  microwave circulator from synthetic rotation},}\ }\href {\doibase
  10.1103/PhysRevApplied.4.034002} {\bibfield  {journal} {\bibinfo  {journal}
  {Phys. Rev. Applied}\ }\textbf {\bibinfo {volume} {4}},\ \bibinfo {pages}
  {034002} (\bibinfo {year} {2015})}\BibitemShut {NoStop}%
\bibitem [{\citenamefont {Onsager}(1931)}]{Onsager1931}%
  \BibitemOpen
  \bibfield  {author} {\bibinfo {author} {\bibfnamefont {L.}~\bibnamefont
  {Onsager}},\ }\bibfield  {title} {\enquote {\bibinfo {title} {Reciprocal
  relations in irreversible processes. i.}}\ }\href {\doibase
  10.1103/PhysRev.37.405} {\bibfield  {journal} {\bibinfo  {journal} {Phys.
  Rev.}\ }\textbf {\bibinfo {volume} {37}},\ \bibinfo {pages} {405--426}
  (\bibinfo {year} {1931})}\BibitemShut {NoStop}%
\bibitem [{\citenamefont {Landau}\ and\ \citenamefont
  {Lifshits}(1980)}]{landau5}%
  \BibitemOpen
  \bibfield  {author} {\bibinfo {author} {\bibfnamefont {L.}~\bibnamefont
  {Landau}}\ and\ \bibinfo {author} {\bibfnamefont {E.}~\bibnamefont
  {Lifshits}},\ }\href {http://books.google.ru/books?id=NaB7oAkon9MC} {\emph
  {\bibinfo {title} {Statistical Physics}}},\ \bibinfo {series} {Course of
  theoretical physics}\ No.\ \bibinfo {number} {pt. 1}\ (\bibinfo  {publisher}
  {Butterworth-Heinemann},\ \bibinfo {year} {1980})\BibitemShut {NoStop}%
\bibitem [{\citenamefont {Estep}\ \emph {et~al.}(2014)\citenamefont {Estep},
  \citenamefont {Sounas}, \citenamefont {Soric},\ and\ \citenamefont
  {Al{\`{u}}}}]{Estep2014}%
  \BibitemOpen
  \bibfield  {author} {\bibinfo {author} {\bibfnamefont {N.~A.}\ \bibnamefont
  {Estep}}, \bibinfo {author} {\bibfnamefont {D.~L.}\ \bibnamefont {Sounas}},
  \bibinfo {author} {\bibfnamefont {J.}~\bibnamefont {Soric}}, \ and\ \bibinfo
  {author} {\bibfnamefont {A.}~\bibnamefont {Al{\`{u}}}},\ }\bibfield  {title}
  {\enquote {\bibinfo {title} {Magnetic-free non-reciprocity and isolation
  based on parametrically modulated coupled-resonator loops},}\ }\href
  {\doibase 10.1038/nphys3134} {\bibfield  {journal} {\bibinfo  {journal}
  {Nature Physics}\ }\textbf {\bibinfo {volume} {10}},\ \bibinfo {pages}
  {923--927} (\bibinfo {year} {2014})}\BibitemShut {NoStop}%
\bibitem [{\citenamefont {Fleury}\ \emph {et~al.}(2014)\citenamefont {Fleury},
  \citenamefont {Sounas}, \citenamefont {Sieck}, \citenamefont {Haberman},\
  and\ \citenamefont {Alu}}]{Fleury2014}%
  \BibitemOpen
  \bibfield  {author} {\bibinfo {author} {\bibfnamefont {R.}~\bibnamefont
  {Fleury}}, \bibinfo {author} {\bibfnamefont {D.~L.}\ \bibnamefont {Sounas}},
  \bibinfo {author} {\bibfnamefont {C.~F.}\ \bibnamefont {Sieck}}, \bibinfo
  {author} {\bibfnamefont {M.~R.}\ \bibnamefont {Haberman}}, \ and\ \bibinfo
  {author} {\bibfnamefont {A.}~\bibnamefont {Alu}},\ }\bibfield  {title}
  {\enquote {\bibinfo {title} {Sound isolation and giant linear nonreciprocity
  in a compact acoustic circulator},}\ }\href {\doibase
  10.1126/science.1246957} {\bibfield  {journal} {\bibinfo  {journal}
  {Science}\ }\textbf {\bibinfo {volume} {343}},\ \bibinfo {pages} {516}
  (\bibinfo {year} {2014})}\BibitemShut {NoStop}%
\bibitem [{\citenamefont {Souslov}\ \emph {et~al.}(2017)\citenamefont
  {Souslov}, \citenamefont {van Zuiden}, \citenamefont {Bartolo},\ and\
  \citenamefont {Vitelli}}]{Souslov2017}%
  \BibitemOpen
  \bibfield  {author} {\bibinfo {author} {\bibfnamefont {A.}~\bibnamefont
  {Souslov}}, \bibinfo {author} {\bibfnamefont {B.~C.}\ \bibnamefont {van
  Zuiden}}, \bibinfo {author} {\bibfnamefont {D.}~\bibnamefont {Bartolo}}, \
  and\ \bibinfo {author} {\bibfnamefont {V.}~\bibnamefont {Vitelli}},\
  }\bibfield  {title} {\enquote {\bibinfo {title} {Topological sound in
  active-liquid metamaterials},}\ }\href {\doibase 10.1038/nphys4193}
  {\bibfield  {journal} {\bibinfo  {journal} {Nature Physics}\ }\textbf
  {\bibinfo {volume} {13}},\ \bibinfo {pages} {1091--1094} (\bibinfo {year}
  {2017})}\BibitemShut {NoStop}%
\bibitem [{\citenamefont {Hafezi}\ and\ \citenamefont
  {Rabl}(2012)}]{Hafezi2012}%
  \BibitemOpen
  \bibfield  {author} {\bibinfo {author} {\bibfnamefont {M.}~\bibnamefont
  {Hafezi}}\ and\ \bibinfo {author} {\bibfnamefont {P.}~\bibnamefont {Rabl}},\
  }\bibfield  {title} {\enquote {\bibinfo {title} {Optomechanically induced
  non-reciprocity in microring resonators},}\ }\href {\doibase
  10.1364/oe.20.007672} {\bibfield  {journal} {\bibinfo  {journal} {Opt.
  Express}\ }\textbf {\bibinfo {volume} {20}},\ \bibinfo {pages} {7672}
  (\bibinfo {year} {2012})}\BibitemShut {NoStop}%
\bibitem [{\citenamefont {Peano}\ \emph {et~al.}(2015)\citenamefont {Peano},
  \citenamefont {Brendel}, \citenamefont {Schmidt},\ and\ \citenamefont
  {Marquardt}}]{Peano2015}%
  \BibitemOpen
  \bibfield  {author} {\bibinfo {author} {\bibfnamefont {V.}~\bibnamefont
  {Peano}}, \bibinfo {author} {\bibfnamefont {C.}~\bibnamefont {Brendel}},
  \bibinfo {author} {\bibfnamefont {M.}~\bibnamefont {Schmidt}}, \ and\
  \bibinfo {author} {\bibfnamefont {F.}~\bibnamefont {Marquardt}},\ }\bibfield
  {title} {\enquote {\bibinfo {title} {Topological phases of sound and
  light},}\ }\href {\doibase 10.1103/physrevx.5.031011} {\bibfield  {journal}
  {\bibinfo  {journal} {Phys. Rev. X}\ }\textbf {\bibinfo {volume} {5}},\
  \bibinfo {pages} {031011} (\bibinfo {year} {2015})}\BibitemShut {NoStop}%
\bibitem [{\citenamefont {Poshakinskiy}\ and\ \citenamefont
  {Poddubny}(2017)}]{Poshakinskiy2017}%
  \BibitemOpen
  \bibfield  {author} {\bibinfo {author} {\bibfnamefont {A.~V.}\ \bibnamefont
  {Poshakinskiy}}\ and\ \bibinfo {author} {\bibfnamefont {A.~N.}\ \bibnamefont
  {Poddubny}},\ }\bibfield  {title} {\enquote {\bibinfo {title} {Phonoritonic
  crystals with a synthetic magnetic field for an acoustic diode},}\ }\href
  {\doibase 10.1103/PhysRevLett.118.156801} {\bibfield  {journal} {\bibinfo
  {journal} {Phys. Rev. Lett.}\ }\textbf {\bibinfo {volume} {118}},\ \bibinfo
  {pages} {156801} (\bibinfo {year} {2017})}\BibitemShut {NoStop}%
\bibitem [{\citenamefont {Brendel}\ \emph {et~al.}(2017)\citenamefont
  {Brendel}, \citenamefont {Peano}, \citenamefont {Painter},\ and\
  \citenamefont {Marquardt}}]{Brendel2017}%
  \BibitemOpen
  \bibfield  {author} {\bibinfo {author} {\bibfnamefont {C.}~\bibnamefont
  {Brendel}}, \bibinfo {author} {\bibfnamefont {V.}~\bibnamefont {Peano}},
  \bibinfo {author} {\bibfnamefont {O.~J.}\ \bibnamefont {Painter}}, \ and\
  \bibinfo {author} {\bibfnamefont {F.}~\bibnamefont {Marquardt}},\ }\bibfield
  {title} {\enquote {\bibinfo {title} {Pseudomagnetic fields for sound at the
  nanoscale},}\ }\href {\doibase 10.1073/pnas.1615503114} {\bibfield  {journal}
  {\bibinfo  {journal} {Proc. Natl. Acad. Sci. USA}\ }\textbf {\bibinfo
  {volume} {114}},\ \bibinfo {pages} {E3390} (\bibinfo {year}
  {2017})}\BibitemShut {NoStop}%
\bibitem [{\citenamefont {Li}\ \emph {et~al.}(2014)\citenamefont {Li},
  \citenamefont {Eggleton}, \citenamefont {Fang},\ and\ \citenamefont
  {Fan}}]{Li2014c}%
  \BibitemOpen
  \bibfield  {author} {\bibinfo {author} {\bibfnamefont {E.}~\bibnamefont
  {Li}}, \bibinfo {author} {\bibfnamefont {B.~J.}\ \bibnamefont {Eggleton}},
  \bibinfo {author} {\bibfnamefont {K.}~\bibnamefont {Fang}}, \ and\ \bibinfo
  {author} {\bibfnamefont {S.}~\bibnamefont {Fan}},\ }\bibfield  {title}
  {\enquote {\bibinfo {title} {Photonic {A}haronov{\textendash}{B}ohm effect in
  photon{\textendash}phonon interactions},}\ }\href {\doibase
  10.1038/ncomms4225} {\bibfield  {journal} {\bibinfo  {journal} {Nat.
  Commun.}\ }\textbf {\bibinfo {volume} {5}},\ \bibinfo {pages} {3225}
  (\bibinfo {year} {2014})}\BibitemShut {NoStop}%
\bibitem [{\citenamefont {Ruesink}\ \emph {et~al.}(2016)\citenamefont
  {Ruesink}, \citenamefont {Miri}, \citenamefont {Al{\`{u}}},\ and\
  \citenamefont {Verhagen}}]{Ruesink2016}%
  \BibitemOpen
  \bibfield  {author} {\bibinfo {author} {\bibfnamefont {F.}~\bibnamefont
  {Ruesink}}, \bibinfo {author} {\bibfnamefont {M.-A.}\ \bibnamefont {Miri}},
  \bibinfo {author} {\bibfnamefont {A.}~\bibnamefont {Al{\`{u}}}}, \ and\
  \bibinfo {author} {\bibfnamefont {E.}~\bibnamefont {Verhagen}},\ }\bibfield
  {title} {\enquote {\bibinfo {title} {Nonreciprocity and magnetic-free
  isolation based on optomechanical interactions},}\ }\href {\doibase
  10.1038/ncomms13662} {\bibfield  {journal} {\bibinfo  {journal} {Nat.
  Commun.}\ }\textbf {\bibinfo {volume} {7}},\ \bibinfo {pages} {13662}
  (\bibinfo {year} {2016})}\BibitemShut {NoStop}%
\bibitem [{\citenamefont {Shen}\ \emph {et~al.}(2018)\citenamefont {Shen},
  \citenamefont {Zhang}, \citenamefont {Chen}, \citenamefont {Sun},
  \citenamefont {Zou}, \citenamefont {Guo}, \citenamefont {Zou},\ and\
  \citenamefont {Dong}}]{Shen2018}%
  \BibitemOpen
  \bibfield  {author} {\bibinfo {author} {\bibfnamefont {Z.}~\bibnamefont
  {Shen}}, \bibinfo {author} {\bibfnamefont {Y.-L.}\ \bibnamefont {Zhang}},
  \bibinfo {author} {\bibfnamefont {Y.}~\bibnamefont {Chen}}, \bibinfo {author}
  {\bibfnamefont {F.-W.}\ \bibnamefont {Sun}}, \bibinfo {author} {\bibfnamefont
  {X.-B.}\ \bibnamefont {Zou}}, \bibinfo {author} {\bibfnamefont {G.-C.}\
  \bibnamefont {Guo}}, \bibinfo {author} {\bibfnamefont {C.-L.}\ \bibnamefont
  {Zou}}, \ and\ \bibinfo {author} {\bibfnamefont {C.-H.}\ \bibnamefont
  {Dong}},\ }\bibfield  {title} {\enquote {\bibinfo {title} {Reconfigurable
  optomechanical circulator and directional amplifier},}\ }\href {\doibase
  10.1038/s41467-018-04187-8} {\bibfield  {journal} {\bibinfo  {journal}
  {Nature Communications}\ }\textbf {\bibinfo {volume} {9}},\ \bibinfo {pages}
  {1797} (\bibinfo {year} {2018})}\BibitemShut {NoStop}%
\bibitem [{\citenamefont {Anguiano}\ \emph {et~al.}(2017)\citenamefont
  {Anguiano}, \citenamefont {Bruchhausen}, \citenamefont {Jusserand},
  \citenamefont {Favero}, \citenamefont {Lamberti}, \citenamefont {Lanco},
  \citenamefont {Sagnes}, \citenamefont {Lema\^{\i}tre}, \citenamefont
  {Lanzillotti-Kimura}, \citenamefont {Senellart},\ and\ \citenamefont
  {Fainstein}}]{Fainstein2017}%
  \BibitemOpen
  \bibfield  {author} {\bibinfo {author} {\bibfnamefont {S.}~\bibnamefont
  {Anguiano}}, \bibinfo {author} {\bibfnamefont {A.~E.}\ \bibnamefont
  {Bruchhausen}}, \bibinfo {author} {\bibfnamefont {B.}~\bibnamefont
  {Jusserand}}, \bibinfo {author} {\bibfnamefont {I.}~\bibnamefont {Favero}},
  \bibinfo {author} {\bibfnamefont {F.~R.}\ \bibnamefont {Lamberti}}, \bibinfo
  {author} {\bibfnamefont {L.}~\bibnamefont {Lanco}}, \bibinfo {author}
  {\bibfnamefont {I.}~\bibnamefont {Sagnes}}, \bibinfo {author} {\bibfnamefont
  {A.}~\bibnamefont {Lema\^{\i}tre}}, \bibinfo {author} {\bibfnamefont {N.~D.}\
  \bibnamefont {Lanzillotti-Kimura}}, \bibinfo {author} {\bibfnamefont
  {P.}~\bibnamefont {Senellart}}, \ and\ \bibinfo {author} {\bibfnamefont
  {A.}~\bibnamefont {Fainstein}},\ }\bibfield  {title} {\enquote {\bibinfo
  {title} {Micropillar resonators for optomechanics in the extremely high
  19--95-{GHz} frequency range},}\ }\href {\doibase
  10.1103/PhysRevLett.118.263901} {\bibfield  {journal} {\bibinfo  {journal}
  {Phys. Rev. Lett.}\ }\textbf {\bibinfo {volume} {118}},\ \bibinfo {pages}
  {263901} (\bibinfo {year} {2017})}\BibitemShut {NoStop}%
\bibitem [{\citenamefont {Fainstein}\ \emph {et~al.}(2013)\citenamefont
  {Fainstein}, \citenamefont {Lanzillotti-Kimura}, \citenamefont {Jusserand},\
  and\ \citenamefont {Perrin}}]{Fainstein2013}%
  \BibitemOpen
  \bibfield  {author} {\bibinfo {author} {\bibfnamefont {A.}~\bibnamefont
  {Fainstein}}, \bibinfo {author} {\bibfnamefont {N.~D.}\ \bibnamefont
  {Lanzillotti-Kimura}}, \bibinfo {author} {\bibfnamefont {B.}~\bibnamefont
  {Jusserand}}, \ and\ \bibinfo {author} {\bibfnamefont {B.}~\bibnamefont
  {Perrin}},\ }\bibfield  {title} {\enquote {\bibinfo {title} {{S}trong
  {O}ptical-{M}echanical {C}oupling in a {V}ertical {G}a{A}s/{A}l{A}s
  {M}icrocavity for {S}ubterahertz {P}honons and {N}ear-{I}nfrared {L}ight},}\
  }\href {\doibase 10.1103/PhysRevLett.110.037403} {\bibfield  {journal}
  {\bibinfo  {journal} {Phys. Rev. Lett.}\ }\textbf {\bibinfo {volume} {110}},\
  \bibinfo {pages} {037403} (\bibinfo {year} {2013})}\BibitemShut {NoStop}%
\bibitem [{\citenamefont {Rozas}\ \emph {et~al.}(2014)\citenamefont {Rozas},
  \citenamefont {Bruchhausen}, \citenamefont {Fainstein}, \citenamefont
  {Jusserand},\ and\ \citenamefont {Lema\^{\i}tre}}]{Rozas2014}%
  \BibitemOpen
  \bibfield  {author} {\bibinfo {author} {\bibfnamefont {G.}~\bibnamefont
  {Rozas}}, \bibinfo {author} {\bibfnamefont {A.~E.}\ \bibnamefont
  {Bruchhausen}}, \bibinfo {author} {\bibfnamefont {A.}~\bibnamefont
  {Fainstein}}, \bibinfo {author} {\bibfnamefont {B.}~\bibnamefont
  {Jusserand}}, \ and\ \bibinfo {author} {\bibfnamefont {A.}~\bibnamefont
  {Lema\^{\i}tre}},\ }\bibfield  {title} {\enquote {\bibinfo {title} {Polariton
  path to fully resonant dispersive coupling in optomechanical resonators},}\
  }\href {\doibase 10.1103/PhysRevB.90.201302} {\bibfield  {journal} {\bibinfo
  {journal} {Phys. Rev. B}\ }\textbf {\bibinfo {volume} {90}},\ \bibinfo
  {pages} {201302} (\bibinfo {year} {2014})}\BibitemShut {NoStop}%
\bibitem [{\citenamefont {Jusserand}\ \emph {et~al.}(2015)\citenamefont
  {Jusserand}, \citenamefont {Poddubny}, \citenamefont {Poshakinskiy},
  \citenamefont {Fainstein},\ and\ \citenamefont {Lemaitre}}]{Jusserand2015}%
  \BibitemOpen
  \bibfield  {author} {\bibinfo {author} {\bibfnamefont {B.}~\bibnamefont
  {Jusserand}}, \bibinfo {author} {\bibfnamefont {A.~N.}\ \bibnamefont
  {Poddubny}}, \bibinfo {author} {\bibfnamefont {A.~V.}\ \bibnamefont
  {Poshakinskiy}}, \bibinfo {author} {\bibfnamefont {A.}~\bibnamefont
  {Fainstein}}, \ and\ \bibinfo {author} {\bibfnamefont {A.}~\bibnamefont
  {Lemaitre}},\ }\bibfield  {title} {\enquote {\bibinfo {title} {Polariton
  resonances for ultrastrong coupling cavity optomechanics in
  $\mathrm{GaAs}/\mathrm{AlAs}$ multiple quantum wells},}\ }\href {\doibase
  10.1103/PhysRevLett.115.267402} {\bibfield  {journal} {\bibinfo  {journal}
  {Phys. Rev. Lett.}\ }\textbf {\bibinfo {volume} {115}},\ \bibinfo {pages}
  {267402} (\bibinfo {year} {2015})}\BibitemShut {NoStop}%
\bibitem [{\citenamefont {Chafatinos}\ \emph {et~al.}(2020)\citenamefont
  {Chafatinos}, \citenamefont {Kuznetsov}, \citenamefont {Anguiano},
  \citenamefont {Bruchhausen}, \citenamefont {Reynoso}, \citenamefont
  {Biermann}, \citenamefont {Santos},\ and\ \citenamefont
  {Fainstein}}]{Chafatinos2020}%
  \BibitemOpen
  \bibfield  {author} {\bibinfo {author} {\bibfnamefont {D.~L.}\ \bibnamefont
  {Chafatinos}}, \bibinfo {author} {\bibfnamefont {A.~S.}\ \bibnamefont
  {Kuznetsov}}, \bibinfo {author} {\bibfnamefont {S.}~\bibnamefont {Anguiano}},
  \bibinfo {author} {\bibfnamefont {A.~E.}\ \bibnamefont {Bruchhausen}},
  \bibinfo {author} {\bibfnamefont {A.~A.}\ \bibnamefont {Reynoso}}, \bibinfo
  {author} {\bibfnamefont {K.}~\bibnamefont {Biermann}}, \bibinfo {author}
  {\bibfnamefont {P.~V.}\ \bibnamefont {Santos}}, \ and\ \bibinfo {author}
  {\bibfnamefont {A.}~\bibnamefont {Fainstein}},\ }\bibfield  {title} {\enquote
  {\bibinfo {title} {Polariton-driven phonon laser},}\ }\href {\doibase
  10.1038/s41467-020-18358-z} {\bibfield  {journal} {\bibinfo  {journal}
  {Nature Communications}\ }\textbf {\bibinfo {volume} {11}},\ \bibinfo {pages}
  {4552} (\bibinfo {year} {2020})}\BibitemShut {NoStop}%
\bibitem [{\citenamefont {Dirac}(1931)}]{Dirac1931}%
  \BibitemOpen
  \bibfield  {author} {\bibinfo {author} {\bibfnamefont {P.~A.~M.}\
  \bibnamefont {Dirac}},\ }\bibfield  {title} {\enquote {\bibinfo {title}
  {Quantised singularities in the electromagnetic field,},}\ }\href {\doibase
  10.1098/rspa.1931.0130} {\bibfield  {journal} {\bibinfo  {journal} {Proc.
  Roy. Soc. London A}\ }\textbf {\bibinfo {volume} {133}},\ \bibinfo {pages}
  {60--72} (\bibinfo {year} {1931})}\BibitemShut {NoStop}%
\bibitem [{\citenamefont {Kavokin}\ \emph {et~al.}(2006)\citenamefont
  {Kavokin}, \citenamefont {Baumberg}, \citenamefont {Malpuech},\ and\
  \citenamefont {Laussy}}]{kavbamalas}%
  \BibitemOpen
  \bibfield  {author} {\bibinfo {author} {\bibfnamefont {A.}~\bibnamefont
  {Kavokin}}, \bibinfo {author} {\bibfnamefont {J.}~\bibnamefont {Baumberg}},
  \bibinfo {author} {\bibfnamefont {G.}~\bibnamefont {Malpuech}}, \ and\
  \bibinfo {author} {\bibfnamefont {F.}~\bibnamefont {Laussy}},\ }\href@noop {}
  {\emph {\bibinfo {title} {{M}icrocavities}}}\ (\bibinfo  {publisher}
  {Clarendon Press},\ \bibinfo {address} {Oxford},\ \bibinfo {year}
  {2006})\BibitemShut {NoStop}%
\bibitem [{\citenamefont {Hanke}\ \emph {et~al.}(1999)\citenamefont {Hanke},
  \citenamefont {Fr\"{o}hlich}, \citenamefont {Ivanov}, \citenamefont
  {Littlewood},\ and\ \citenamefont {Stolz}}]{Hanke1999}%
  \BibitemOpen
  \bibfield  {author} {\bibinfo {author} {\bibfnamefont {L.}~\bibnamefont
  {Hanke}}, \bibinfo {author} {\bibfnamefont {D.}~\bibnamefont {Fr\"{o}hlich}},
  \bibinfo {author} {\bibfnamefont {A.~L.}\ \bibnamefont {Ivanov}}, \bibinfo
  {author} {\bibfnamefont {P.~B.}\ \bibnamefont {Littlewood}}, \ and\ \bibinfo
  {author} {\bibfnamefont {H.}~\bibnamefont {Stolz}},\ }\bibfield  {title}
  {\enquote {\bibinfo {title} {{LA} phonoritons in {C}u$_2${O}},}\ }\href
  {\doibase 10.1103/physrevlett.83.4365} {\bibfield  {journal} {\bibinfo
  {journal} {Phys. Rev. Lett.}\ }\textbf {\bibinfo {volume} {83}},\ \bibinfo
  {pages} {4365} (\bibinfo {year} {1999})}\BibitemShut {NoStop}%
\bibitem [{\citenamefont {{Keldysh}}\ and\ \citenamefont
  {{Tikhodeev}}(1986)}]{Keldysh1986}%
  \BibitemOpen
  \bibfield  {author} {\bibinfo {author} {\bibfnamefont {L.~V.}\ \bibnamefont
  {{Keldysh}}}\ and\ \bibinfo {author} {\bibfnamefont {S.~G.}\ \bibnamefont
  {{Tikhodeev}}},\ }\bibfield  {title} {\enquote {\bibinfo {title}
  {{High-intensity polariton wave near the stimulated scattering threshold}},}\
  }\href@noop {} {\bibfield  {journal} {\bibinfo  {journal} {Zh. Eksp. Teor.
  Fiz.}\ }\textbf {\bibinfo {volume} {90}},\ \bibinfo {pages} {1852--1870}
  (\bibinfo {year} {1986})},\ \translation{Sov. Phys. JETP {\bf 63}, 1086
  (1986)}\BibitemShut {NoStop}%
\bibitem [{\citenamefont {Ivanov}\ and\ \citenamefont
  {{Keldysh}}(1982)}]{Ivanov1982}%
  \BibitemOpen
  \bibfield  {author} {\bibinfo {author} {\bibfnamefont {A.~L.}\ \bibnamefont
  {Ivanov}}\ and\ \bibinfo {author} {\bibfnamefont {L.}~\bibnamefont
  {{Keldysh}}},\ }\bibfield  {title} {\enquote {\bibinfo {title} {Restructuring
  of polariton and phonon spectra of a semiconductor in the presence of a
  strong electromagnetic wave},}\ }\href@noop {} {\bibfield  {journal}
  {\bibinfo  {journal} {Zh. Eksp. Teor. Fiz.}\ }\textbf {\bibinfo {volume}
  {84}},\ \bibinfo {pages} {404} (\bibinfo {year} {1982})},\ \translation{Sov.
  Phys. JETP {\bf 57}, 234 (1983)}\BibitemShut {NoStop}%
\bibitem [{\citenamefont {Baker}\ \emph {et~al.}(2014)\citenamefont {Baker},
  \citenamefont {Hease}, \citenamefont {Nguyen}, \citenamefont {Andronico},
  \citenamefont {Ducci}, \citenamefont {Leo},\ and\ \citenamefont
  {Favero}}]{Baker2014}%
  \BibitemOpen
  \bibfield  {author} {\bibinfo {author} {\bibfnamefont {C.}~\bibnamefont
  {Baker}}, \bibinfo {author} {\bibfnamefont {W.}~\bibnamefont {Hease}},
  \bibinfo {author} {\bibfnamefont {D.-T.}\ \bibnamefont {Nguyen}}, \bibinfo
  {author} {\bibfnamefont {A.}~\bibnamefont {Andronico}}, \bibinfo {author}
  {\bibfnamefont {S.}~\bibnamefont {Ducci}}, \bibinfo {author} {\bibfnamefont
  {G.}~\bibnamefont {Leo}}, \ and\ \bibinfo {author} {\bibfnamefont
  {I.}~\bibnamefont {Favero}},\ }\bibfield  {title} {\enquote {\bibinfo {title}
  {Photoelastic coupling in gallium arsenide optomechanical disk resonators},}\
  }\href {\doibase 10.1364/OE.22.014072} {\bibfield  {journal} {\bibinfo
  {journal} {Opt. Express}\ }\textbf {\bibinfo {volume} {22}},\ \bibinfo
  {pages} {14072--14086} (\bibinfo {year} {2014})}\BibitemShut {NoStop}%
\bibitem [{\citenamefont {Aspelmeyer}\ \emph {et~al.}(2014)\citenamefont
  {Aspelmeyer}, \citenamefont {Kippenberg},\ and\ \citenamefont
  {Marquardt}}]{Kippenberg2014}%
  \BibitemOpen
  \bibfield  {author} {\bibinfo {author} {\bibfnamefont {M.}~\bibnamefont
  {Aspelmeyer}}, \bibinfo {author} {\bibfnamefont {T.~J.}\ \bibnamefont
  {Kippenberg}}, \ and\ \bibinfo {author} {\bibfnamefont {F.}~\bibnamefont
  {Marquardt}},\ }\bibfield  {title} {\enquote {\bibinfo {title} {Cavity
  optomechanics},}\ }\href {\doibase 10.1103/RevModPhys.86.1391} {\bibfield
  {journal} {\bibinfo  {journal} {Rev. Mod. Phys.}\ }\textbf {\bibinfo {volume}
  {86}},\ \bibinfo {pages} {1391} (\bibinfo {year} {2014})}\BibitemShut
  {NoStop}%
\bibitem [{\citenamefont {Berry}(1984)}]{Berry1984}%
  \BibitemOpen
  \bibfield  {author} {\bibinfo {author} {\bibfnamefont {M.~V.}\ \bibnamefont
  {Berry}},\ }\bibfield  {title} {\enquote {\bibinfo {title} {{Q}uantal {P}hase
  {F}actors {A}ccompanying {A}diabatic {C}hanges},}\ }\href {\doibase
  10.1098/rspa.1984.0023} {\bibfield  {journal} {\bibinfo  {journal} {Proc.
  Roy. Soc. London A}\ }\textbf {\bibinfo {volume} {392}},\ \bibinfo {pages}
  {45--57} (\bibinfo {year} {1984})}\BibitemShut {NoStop}%
\bibitem [{\citenamefont {Poshakinskiy}\ \emph {et~al.}(2016)\citenamefont
  {Poshakinskiy}, \citenamefont {Poddubny},\ and\ \citenamefont
  {Fainstein}}]{Poshakinskiy2016PRL}%
  \BibitemOpen
  \bibfield  {author} {\bibinfo {author} {\bibfnamefont {A.~V.}\ \bibnamefont
  {Poshakinskiy}}, \bibinfo {author} {\bibfnamefont {A.~N.}\ \bibnamefont
  {Poddubny}}, \ and\ \bibinfo {author} {\bibfnamefont {A.}~\bibnamefont
  {Fainstein}},\ }\bibfield  {title} {\enquote {\bibinfo {title} {Multiple
  quantum wells for $\mathcal{P}\mathcal{T}$-symmetric phononic crystals},}\
  }\href {\doibase 10.1103/PhysRevLett.117.224302} {\bibfield  {journal}
  {\bibinfo  {journal} {Phys. Rev. Lett.}\ }\textbf {\bibinfo {volume} {117}},\
  \bibinfo {pages} {224302} (\bibinfo {year} {2016})}\BibitemShut {NoStop}%
\bibitem [{\citenamefont {Bahari}\ \emph {et~al.}(2017)\citenamefont {Bahari},
  \citenamefont {Ndao}, \citenamefont {Vallini}, \citenamefont {El~Amili},
  \citenamefont {Fainman},\ and\ \citenamefont {Kant{\'e}}}]{Bahari2017}%
  \BibitemOpen
  \bibfield  {author} {\bibinfo {author} {\bibfnamefont {B.}~\bibnamefont
  {Bahari}}, \bibinfo {author} {\bibfnamefont {A.}~\bibnamefont {Ndao}},
  \bibinfo {author} {\bibfnamefont {F.}~\bibnamefont {Vallini}}, \bibinfo
  {author} {\bibfnamefont {A.}~\bibnamefont {El~Amili}}, \bibinfo {author}
  {\bibfnamefont {Y.}~\bibnamefont {Fainman}}, \ and\ \bibinfo {author}
  {\bibfnamefont {B.}~\bibnamefont {Kant{\'e}}},\ }\bibfield  {title} {\enquote
  {\bibinfo {title} {Nonreciprocal lasing in topological cavities of arbitrary
  geometries},}\ }\href {\doibase 10.1126/science.aao4551} {\bibfield
  {journal} {\bibinfo  {journal} {Science}\ }\textbf {\bibinfo {volume}
  {358}},\ \bibinfo {pages} {636} (\bibinfo {year} {2017})}\BibitemShut
  {NoStop}%
\bibitem [{\citenamefont {Bandres}\ \emph {et~al.}(2018)\citenamefont
  {Bandres}, \citenamefont {Wittek}, \citenamefont {Harari}, \citenamefont
  {Parto}, \citenamefont {Ren}, \citenamefont {Segev}, \citenamefont
  {Christodoulides},\ and\ \citenamefont {Khajavikhan}}]{Segev2018b}%
  \BibitemOpen
  \bibfield  {author} {\bibinfo {author} {\bibfnamefont {M.~A.}\ \bibnamefont
  {Bandres}}, \bibinfo {author} {\bibfnamefont {S.}~\bibnamefont {Wittek}},
  \bibinfo {author} {\bibfnamefont {G.}~\bibnamefont {Harari}}, \bibinfo
  {author} {\bibfnamefont {M.}~\bibnamefont {Parto}}, \bibinfo {author}
  {\bibfnamefont {J.}~\bibnamefont {Ren}}, \bibinfo {author} {\bibfnamefont
  {M.}~\bibnamefont {Segev}}, \bibinfo {author} {\bibfnamefont {D.~N.}\
  \bibnamefont {Christodoulides}}, \ and\ \bibinfo {author} {\bibfnamefont
  {M.}~\bibnamefont {Khajavikhan}},\ }\bibfield  {title} {\enquote {\bibinfo
  {title} {Topological insulator laser: Experiments},}\ }\href {\doibase
  10.1126/science.aar4005} {\bibfield  {journal} {\bibinfo  {journal}
  {Science}\ }\textbf {\bibinfo {volume} {359}},\ \bibinfo {pages} {eaar4005}
  (\bibinfo {year} {2018})}\BibitemShut {NoStop}%
\bibitem [{\citenamefont {Babikov}(1966)}]{babikov1966method}%
  \BibitemOpen
  \bibfield  {author} {\bibinfo {author} {\bibfnamefont {V.}~\bibnamefont
  {Babikov}},\ }\href@noop {} {\emph {\bibinfo {title} {Method of Phase
  Functions in Quantum Mechanics}}},\ \bibinfo {type} {Tech. Rep.}\ (\bibinfo
  {institution} {Joint Inst. for Nuclear Research, Dubna (USSR). Lab. of
  Theoretical Physics},\ \bibinfo {year} {1966})\BibitemShut {NoStop}%
\bibitem [{\citenamefont {Denisov}\ \emph {et~al.}(2017)\citenamefont
  {Denisov}, \citenamefont {Rozhansky}, \citenamefont {Averkiev},\ and\
  \citenamefont {L\"{a}hderanta}}]{Denisov2017}%
  \BibitemOpen
  \bibfield  {author} {\bibinfo {author} {\bibfnamefont {K.~S.}\ \bibnamefont
  {Denisov}}, \bibinfo {author} {\bibfnamefont {I.~V.}\ \bibnamefont
  {Rozhansky}}, \bibinfo {author} {\bibfnamefont {N.~S.}\ \bibnamefont
  {Averkiev}}, \ and\ \bibinfo {author} {\bibfnamefont {E.}~\bibnamefont
  {L\"{a}hderanta}},\ }\bibfield  {title} {\enquote {\bibinfo {title} {A
  nontrivial crossover in topological {H}all effect regimes},}\ }\href
  {\doibase 10.1038/s41598-017-16538-4} {\bibfield  {journal} {\bibinfo
  {journal} {Scientific Reports}\ }\textbf {\bibinfo {volume} {7}},\ \bibinfo
  {pages} {17204} (\bibinfo {year} {2017})}\BibitemShut {NoStop}%
\end{thebibliography}

%

\end{document}